\documentclass[traditabstract]{aa}
\usepackage{graphicx}
\usepackage{amsmath}
\usepackage{amssymb}
\usepackage{bbm}
\usepackage{natbib}
\usepackage{epstopdf}
\usepackage{ulem}
\usepackage{hyperref}

\hypersetup{colorlinks,breaklinks,
            urlcolor=[rgb]{0,0,0.5},
            linkcolor=[rgb]{0,0,0.5},
            citecolor=[rgb]{0,0,0.5}}

\usepackage{layouts}
\usepackage{color}
    \definecolor{Blue}{rgb}{0.0,0.0,1.0}
    \definecolor{Red}{rgb}{1.0,0.0,0.0}
    \definecolor{Green}{rgb}{0.0,1.0,0.0}

\begin{document}
\title
{
Limits on thickness and efficiency of Polish doughnuts in application to the ULX sources}
   \author{M. Wielgus\inst{1,2}
     \and
          W. Yan\inst{1}
     \and
         J.-P. Lasota\inst{1,3}
     \and
          M. A. Abramowicz\inst{1,4,5}
          }
          
          \titlerunning{Polish doughnuts in application to the ULXs}
   \institute{ N. Copernicus Astronomical Center, ul. Bartycka 18, PL-00-716 Warszawa, Poland \\
              \email{maciek.wielgus@gmail.com ; wyan@camk.edu.pl}
               \and
             Institute of Micromechanics and Photonics, Warsaw University of Technology, ul. \'Sw. A. Boboli 8,\\ PL-02-525 Warszawa, Poland
     \and
             Institut d'Astrophysique de Paris, CNRS et Sorbonne Universit\'es, UPMC
             Univ Paris~06, UMR 7095,  \\ 98bis Bd Arago, 75014 Paris, France \\
             \email{lasota@iap.fr}
     \and
             Physics Department, Gothenburg University, SE-412-96 G{\"o}teborg, Sweden\thanks{Professor Emeritus} \\
             \email{marek.abramowicz@physics.gu.se}
     \and
             Institute of Physics, Silesian University in Opava, Bezru{\v c}ovo n{\'a}m. 13, CZ-746 01, Opava, Czech Republic
             }
   \date{Received December, 2015; accepted December 2015}
\abstract
{Polish doughnuts (PDs) are geometrically thick disks that rotate with super-Keplerian velocities in their innermost parts, and whose long and narrow funnels along rotation axes collimate the emerging radiation into beams. In this paper we construct an extremal family of PDs that maximize both geometrical thickness and radiative efficiency. We then derive upper limits for these quantities and subsequently for the related ability to collimate radiation.
PDs with such extreme properties may explain the observed properties of ultraluminous X-ray sources without the need for the black hole masses to exceed $\sim 10 M_\odot$. However, we show that strong advective cooling, which is expected to be one of the dominant cooling mechanisms in accretion flows with super-Eddington accretion rates, tends to reduce the geometrical thickness and luminosity of PDs substantially. We also show that the beamed radiation emerging from the PD funnels corresponds to ``isotropic'' luminosities that obey $L_{\rm col} \approx 0.1 {\dot M}c^2$ for ${\dot M} \gg {\dot M}_{\rm Edd}$, and not the familiar and well-known logarithmic relation, $L \sim \ln {\dot M}$. }
   \maketitle
%
%
%

\section{Introduction}

The research reported here was motivated by the question whether collimation of radiation in the funnels of Polish doughnuts (PDs) can explain super-Eddington luminosities of the ultraluminous X-ray (ULX) sources, assuming that the ULXs are powered by accretion on the stellar mass compact objects.

\subsection{Super-Eddington accretion}
%
The Eddington luminosity for an object with a mass $M$ is given by the formula,
\begin{equation}
L_{\rm Edd} \equiv \frac{4\pi G M m_p c}{\sigma_T}
= 1.3 \times 10^{38} \left( \frac{M}{M_\odot} \right) {\rm[erg/sec].}
\label{Eddington-luminosity}
\end{equation}
Here $\sigma_T$ is the electron scattering cross-section and $m_p$ is the proton mass. For objects powered by accretion, the corresponding Eddington accretion rate ${\dot M}_{\rm Edd}$ is defined by\footnotemark
\begin{equation}
{\dot M}_{\rm Edd} \equiv \frac{L_{\rm Edd}}{0.1 c^2}
= 1.4 \times 10^{18} \left( \frac{M}{M_\odot} \right) {\rm[g/sec]}.
\end{equation}
%
\footnotetext{We caution that several authors used different definitions of the Eddington accretion rate, ${\dot M}_{\rm Edd} =
{L_{\rm Edd}}/\eta {c^2}$, with $\eta$ being the the efficiency of accretion. $\eta = 0.1$ is most frequently adopted, but some authors used $\eta = 1/16$ or some other values.}
Observations provide several examples of objects that radiate at super-Eddington luminosities. In our Galaxy the best-known
examples are SS433 and GRS 1915+105 \citep[see e.g.][and references therein]{Fabrika-2006, Fender-2004}. Outside the Galaxy, super-Eddington luminosities are reached by ULX sources and tidal disruption events
\citep[e.g.][]{Fabbiano-2006, vanVelzen-2014}, as well as by several AGNs \citep[see e.g.][and references therein]{Du-2015}.
In the aspects that are relevant to our paper, the theory of super-Eddington accretion onto black holes was reviewed by
\cite{Paczynski-1982}, \cite{Paczynski-1998}, or \cite{Abramowicz-2005}.

It is convenient to introduce the
rescaled luminosity $\lambda$,
rescaled accretion rate ${\dot m}$,
maximal relative vertical thickness  $\chi$,
and dimensionless advection strength $\xi$,
\begin{equation}
\lambda \equiv \frac{L}{L_{\rm Edd}}, ~~{\dot m} \equiv \frac{{\dot M}}{{\dot M}_{\rm Edd}}, ~~ \chi = \left(
\frac{H}{R}\right)_{\rm max}, \\
\label{Eddington-rescaled}
\end{equation}
\begin{equation}
\xi = \frac{L_{\rm adv}}{L_0} = \frac{\rm advective~energy~losses}{\rm total~energy~generation}, ~~~~
0 \le \xi \le 1.
\label{advective-cooling-parameter}
\end{equation}
Here $H=H(R)$ is the vertical semi-thickness of the disk at the distance $R$ from the black hole. 
%
%
%
\subsection{Polish doughnuts and the ULX sources}
%
   \begin{figure}[h]
  \includegraphics[width=0.5\textwidth, trim = {0mm 0 0 5}, clip]{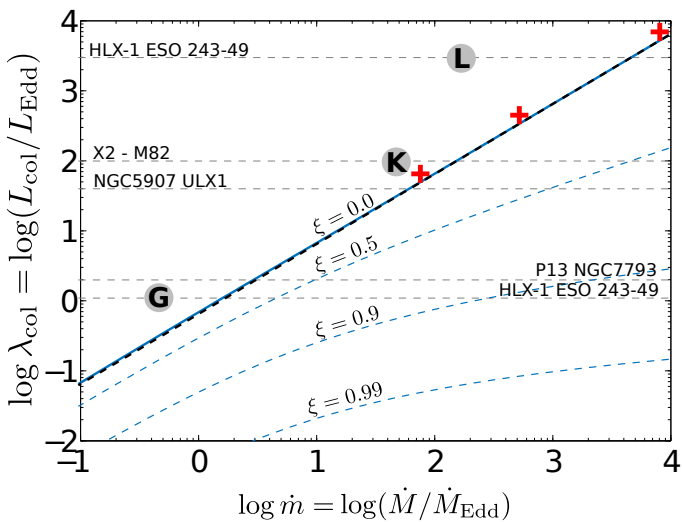}
\caption{
Observed luminosities of a few ULX sources (horizontal dashed lines) compared with our extremal Polish doughnuts (blue lines) for different strength of advection $\xi$. We also indicate results of the global
MHD simulations (red crosses) and various ad hoc models from recent literature (gray circles with letters). See text for a~detailed explanation.
}\label{Figure-main-results}

   \end{figure}

The main results of our paper are summarized in Fig. \ref{Figure-main-results}, where luminosities inferred from different models are plotted against the corresponding mass accretion rates. Results related to the {\sl extremal} PDs, discussed in Sect. \ref{sec:maximalThick}, are indicated with blue lines. Different curves correspond to different advection strengths, expressed by the parameter $\xi$, as defined by Eq. (\ref{advective-cooling-parameter}). The solid blue line indicates $\xi =0.0$ (no advection), the other three blue dashed lines are for $\xi = 0.5, 0.9,$ and $0.99$.
The collimated luminosity, $L_{\rm col} \approx  L/(1 - \cos \alpha)$, emerging from a~narrow funnel with opening half-angle $\alpha$, is an isotropic equivalent luminosity, that may be interpreted as the
observed luminosity of sources in which radiation is strongly beamed (cf. Subsect. \ref{ssec:L}).
The theoretical limit $\lambda_{\rm col} = L_{\rm col}/L_{\rm Edd} = 0.625 \dot{m}$, derived in Sect. \ref{section-efficiency-limit}, is denoted with a~black dashed line in Fig. \ref{Figure-main-results}. It approximates the luminosity of a~non-advective torus (thick blue continuous line) very accurately. Only this advectionless maximal configuration is close to points representing ULXs; accounting for advection leads to far too dim sources.
The thin horizontal lines (with labels) correspond to the observed X-ray luminosities of ULX sources. Two with known masses,
a neutron star X-2 in M\,82 with mass $M\approx 1.4 M_\odot$ and luminosity $L_X = 1.8 \times 10^{40}$ erg/s,
\citep{Bachetti-2014} and a black hole NGC\,7793 with mass $M < 15 M_\odot$ and $L_X \approx 5 \times 10^{39}$ erg/s,
\citep{Motch-2014}, and two ULX sources with unknown or controversial masses, HLX-1 in ESO 243-49 with
$L_X=1.2 \times 10^{42}$ erg/s \citep{Farrell-2009, Godet-2010}, and NGC\,5907 ULX1 with $L_X \approx 5\times 10^{40}$ erg/s
\citep{Walton-2015}. For the last source we assumed a~mass of $M = 10M_\odot$, for HLX-1 we considered the cases of two proposed masses: $10^4M_\odot$ and $3M_\odot$. Circles on these lines show locations of the theoretical models, proposed by \cite{Godet-2012}(G), \cite{Kluzniak-2015} (K), and \cite{Lasota-2015ApJL} (L).
The three crosses correspond to three models of black hole accretion flows from a recent magnetohydrodynamical (MHD) numerical simulation by
\cite{Sadowski-2015}. These simulations have been done assuming $M = 3\times 10^5 M_\odot$, but in the $L/L_{\rm Edd}$ versus
${\dot M}/{\dot M}_{\rm Edd}$ relation, dependencies on the mass are scaled off.

\cite{Lasota-2015p} showed that slim accretion disks in which cooling is dominated by advection
\citep{Abramowicz-1988, Sadowski-2009} cannot be very geometrically thick, and that even at highly super-Eddington accretion
rates, ${\dot m} \gg 1$, the maximal relative disk thickness stays rather small, $\cot \alpha = \chi \le 5$. Thus, the
disks are indeed slim.  On the other hand, models of PDs are well-known to have arbitrarily large thickness,
$\chi \gg 5$ \citep{Abramowicz-1978, Kozlowski-1978, Paczynski-1980, Jaroszynski-1980}. Here, we resolve this apparent
contradiction by pointing out that the models of PDs constructed so far have been non-advective.
%
%
We reconsidered the PD models to include a strong, global advective cooling. We show that taking advection into account greatly reduces the PD thickness, deeming very thick tori construction impossible for
realistic mass accretion rates. We proceed by evaluating the magnitude of radiation collimation in a funnel of a very thick PD
\citep{Sikora-1981}, finding that it obeys the linear scaling $L_{\rm col} = 0.0625\dot{M} c^2$, which agrees reasonably well with recent numerical simulations \citep{Sadowski-2015} and observations of ULXs. However, when advection is accounted for, PDs cannot provide sufficient luminosity to explain ULXs.

\begin{figure*}
   \centering
  \includegraphics[width=0.90\textwidth]{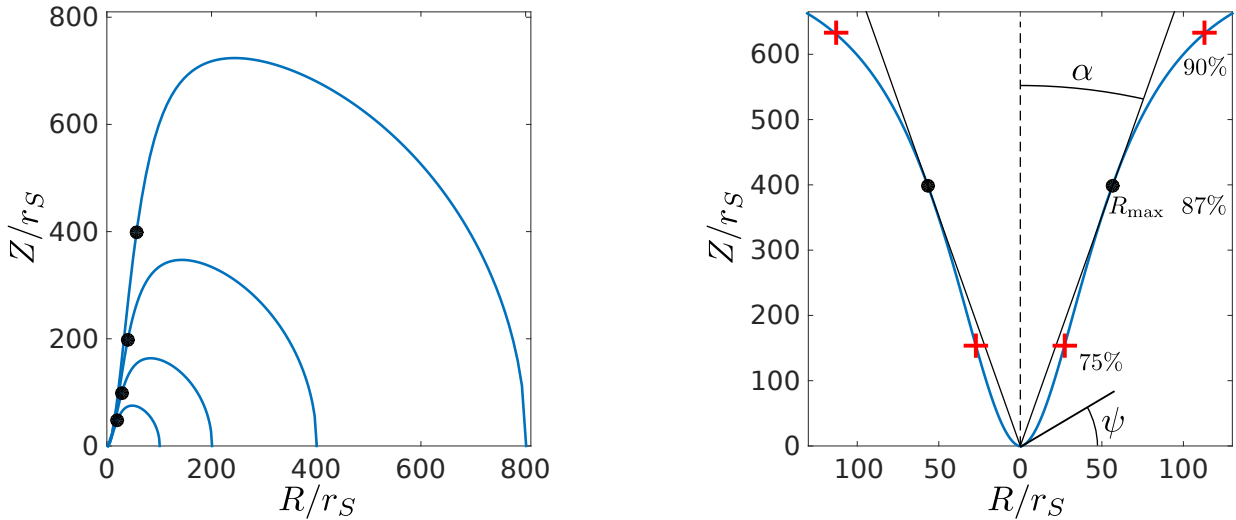}
   \caption{
   {\it Left}: Shapes of tori with constant angular momentum. PDs with
   $R_{\rm in} \rightarrow R_{\rm mb}$ have an arbitrarily high $H/R$, see also Fig. \ref{Figure-constant-H-R-max}.
   Shapes correspond to $R_{\rm in}/r_{\mathrm S} = 2.025, 2.01, 2.005,$ and $2.0025$. Black dots indicate the highest $H/R$.
   {\it Right}: Torus with $R_{\rm in}/r_{\mathrm S} = 2.0025$, corresponding to $\chi =  \cot \alpha = 7.04$. The dot indicates the highest $H/R$, crosses indicate the outer
   boundary of the region, where 75\% and 90\% of the total radiation is generated. We note that the scaling is different for vertical and horizontal axes.
   }
   \label{Figure-constant-shape}
\end{figure*}

\section{Polish doughnuts: assumptions and equations}


Polish doughnuts  are stationary and axially symmetric models of accretion structures around black holes. All properties of a PD 
are derived from a~single assumed function $\ell(R)$: the specific angular momentum distribution at the PD surface. From an assumed $\ell(R)$ the PD shape $H(R)$ is calculated together with the radiation flux at the surface $\vec{f}(R)$, the total luminosity $L = \int \vec{f} d\vec{S},$ and finally the accretion rate ${\dot M}$. All these are given in terms of analytic
(algebraic) formulae. No physical properties of the PD interior need to be considered. We stress that models of PDs do not assume anything specific about their interiors, not even about the equation of state, $p = p(\rho,
T)$. In particular, the pressure (gas and radiation) $p$, the density $\rho$, the temperature $T,$ and the (non-azimuthal)
velocity $\vec{v}$ do not appear in the model.
\par
The PDs were constructed using the Einstein relativistic hydrodynamics equations in the Kerr
geometry ($\nabla_\mu T^\mu_{~\nu} = 0$, etc.), but they are often considered in the Newtonian model of the gravity of a~non-rotating
black hole introduced by Bohdan Paczy{\'n}ski. The Paczy{\'n}ski model assumes Newtonian hydrodynamics and the gravitational
potential given by the \cite{Paczynski-1980} formula,
\begin{equation}
\Phi = - \frac{GM}{r - r_{\mathrm S}}, ~~~ r = (R^2 + Z^2)^{1/2}, ~~~ r_{\mathrm S} = \frac{2GM}{c^2},
\label{Paczynski-Wiita-potential}
\end{equation}
where $R$, $Z$, $\phi$ are cylindrical coordinates, $M$ is the black hole mass, and ${r_{\mathrm S}}$ is the gravitational {(Schwarzschild)}  radius.
In the Paczy{\'n}ski potential, the marginally stable orbit (ISCO) is located at $R_{\rm ms} = 3{r_{\mathrm S}}$ and the marginally
bound circular orbit at $R_{\rm mb} = 2{r_{\mathrm S}}$, exactly as in the case of the Schwarzschild (non-rotating) black hole.
\vskip0.1truecm \noindent                                                               
The ``classic'' PD models assume that
\begin{enumerate}
\item the photosphere coincides with an equipressure surface,
\begin{equation}
p = p(R, Z) = p_0 = \mbox{const, and that}
\label{location-photosphere}
\end{equation}
\item the specific angular momentum $\ell$ at the photosphere is a~known (assumed) function,
\begin{equation}
\ell = \ell(R) .
\label{angular-momentum-at-photosphere}
\end{equation}
The angular momentum is Keplerian at the inner and outer radii of a PD, $\ell(R_{\rm in}) = \ell_K(R_{\rm in})$ and $\ell(R_{\rm
out}) = \ell_K(R_{\rm out})$, where $R_{\rm mb} < R_{\rm in} < R_{\rm ms}$.
\item Radiation is emitted from the photosphere at the local Eddington rate, that is, the local flux $\vec{f}^{\rm rad}$ is  given by
\begin{equation}
\vec{f}^{\rm rad} = \frac{c}{\kappa}\vec{g}^{\rm eff} ,
\label{locally-Eddington}
\end{equation}
where $\kappa$ is the mass absorption coefficient and $\vec{g}^{\rm eff}$ is the effective gravity, given by
\begin{eqnarray}
 g^{\rm eff}_R  \equiv \frac{1}{\rho} \frac{\partial p}{\partial R} &=& -\frac{\partial \Phi}{\partial R} + \frac{\ell^2(R)}{R^3},
\label{balance-forces-R} \\
g^{\rm eff}_Z \equiv \frac{1}{\rho} \frac{\partial p}{\partial Z} &=& -\frac{\partial \Phi}{\partial Z} .
\label{balance-forces-Z}
\end{eqnarray}
\end{enumerate}
To parametrize PD solutions we use the dimensionless parameter $\rho$,
\begin{equation}
\rho = \frac{R_{\rm in} - R_{\rm mb}}{r_{\mathrm S}} , \ 0 \leq \rho \leq 1 .
\label{eq:rho}
\end{equation}

\subsection{Polish doughnut shape $H(R)$}
By integrating the differential equation
\begin{equation}
\frac{dH}{dR} = - \left(\frac{g^{\rm eff}_R}{g^{\rm eff}_Z}\right)_{Z = H} \equiv F(R, H),
\label{integration-shape}
\end{equation}
we derive an explicit analytic formula for the PD shape, $H = H(R)$,
\begin{equation} \label{shape-photosphere-01}
\begin{split}
&H(R) = \left\{ \left[ \frac{GM ( R_{\rm in} - r_{\rm S}) }{GM - ( R_{\rm in} - r_{\mathrm S})I(R)} +
r_{\rm S} \right]^2 - R^2 \right\}^{1/2}, \\
&{\rm{where}}~~~I(R) \equiv \int^R_{R_{\rm in}}\ell^2(R')R'^{-3} d R'.
\end{split}
\end{equation}
The vertical thickness $H(R)$ is zero at the inner edge,
$R_{\rm in}$.
It is easy to see that the thickness is also zero at the outer edge $R_{\rm out}$ given by the integral condition,
\begin{equation}
\int^{R_{\rm out}}_{R_{\rm in}} \left[ \frac{\ell_{\mathrm K}^2(R) - \ell^2(R)}{R^3} \right] dR = 0.
\label{integral-condition}
\end{equation}
In the particular case of a~constant angular momentum distribution,
\begin{equation}
\ell = \ell_{\rm K}(R_{\rm in}) = \left[  \frac{G\,M\,R_{\rm in}^3}{(R_{\rm in} -r_{\mathrm S})^2}   \right]^{1/2} = {\rm const},
\label{constant-angular-momentum}
\end{equation}
 from Eq. (\ref{integral-condition}) we obtain
\begin{equation}
R_{\rm out} = \frac{R_{\rm in}r_{\mathrm S}}{R_{\rm in} - 2r_{\mathrm S}} .
\label{constant-outer-edge}
\end{equation}
Then the shape of the constant angular momentum torus is
\begin{equation}
H(R) =    \left\{ \left[  \frac{2(R_{\rm in} - r_{\mathrm S})^2 R^2}{R^2(R_{\rm in} - 2 r_{\mathrm S}) + R_{\rm in}^3}{+} r_{\mathrm S} \right]^2 {-} R^2 \right\}^{1/2} .
\label{constant-shape}
\end{equation}
Two limiting shapes are the infinite, unbounded torus for $R_{\rm in} = 2 r_{\mathrm S}$ ($\rho = 0$) and the ring whose cross section is reduced to a~point for $R_{\rm in} = 3 r_{\mathrm S}$ ($\rho = 1$). Examples of $H(R)$ profiles for the constant angular momentum distribution, calculated assuming different $R_{\rm in}$, are shown in Fig. \ref{Figure-constant-shape} (left). 
\subsection{Polish doughnut luminosity $L$}
\label{ssec:L}
Integrating $\vec{f}^{\rm rad}$ over the PD photosphere, whose location is given by Eq. (\ref{shape-photosphere-01}), gives
the PD total luminosity $L$,
\begin{align}
\begin{split}
&\frac{L}{L_{\rm Edd}} = \int\limits^{R_{\rm out}}_{R_{\rm in}} \left[ \frac{\ell^4 r (r - r_{\mathrm S})^2}{G^2 M^2 H R^5} - \frac{2\ell^2}{GM H R} + \frac{r
R}{H(r - r_{\mathrm S})^2} \right] dR ,\\
&{\rm{where}} ~~~ r \equiv (R^2 + H^2)^{1/2}.
\label{luminosity-integration-result}
\end{split}
\end{align}
For thick tori most contribution to the integral (\ref{luminosity-integration-result}) comes from the inner region, that is, the funnel, $L_{\rm fun} \approx L$, Fig. \ref{Figure-constant-shape} (right). When radiation collimated by the funnel of opening half-angle $\alpha$ is observed, which yields a~measured flux $f^r$, the collimated equivalent isotropic luminosity is calculated to be
\begin{equation}
L_{\rm col} = 4 \pi r^2 f^r = \frac{L_{\rm fun}}{1-\cos \alpha} = \beta L_{\rm fun} \approx \beta L , 
\end{equation}
where we have introduced the collimating factor $\beta$.
%
   \begin{figure}
   \centering
   \includegraphics[width=0.48\textwidth]{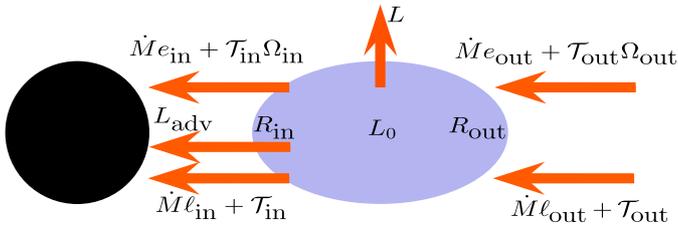}
   \caption{Conservation of energy and angular momentum}
   \label{conservation-energy-momentum}
   \end{figure}
%
\subsection{Polish doughnut efficiency $\epsilon$}
Figure \ref{conservation-energy-momentum} illustrates the conservation of energy and angular momentum in a PD. Energy
and angular momentum flow in at the outer edge and flow out at the inner edge.  The rate at which energy is deposited in the
PD interior is $L_{0}$,
\begin{align}
& ({\dot M}e_{\rm out} + {\cal T}_{\rm out}\Omega_{\rm out}) - ({\dot M}e_{\rm in} + {\cal T}_{\rm in}\Omega_{\rm in}) = L_{
0}, \label{conservation-energy}
\\
& ({\dot M}\ell_{\rm out} + {\cal T}_{\rm out}) - ({\dot M}\ell_{\rm in} + {\cal T}_{\rm in}) = 0.
\label{conservation-momentum}
\end{align}
Here ${\dot M}$ is the accretion rate,
\begin{equation}
e = \Phi + \frac{\ell^2}{2 R^2}
\label{mechanical-energy}
\end{equation}
is the specific mechanical energy, $\Omega = \ell/R^2$ is the angular velocity, and ${\cal T}$ is the torque. Assuming the usual
no-torque inner boundary condition, ${\cal T}_{\rm in} = 0$, we derive from Eqs.
(\ref{conservation-energy})-(\ref{conservation-momentum})
\begin{equation}
L_{0} =  - \epsilon(R_{\rm in}, R_{\rm out}) \, {\dot M} ,
\label{energy-generation-interior}
\end{equation}
where $\epsilon(R_{\rm in}, R_{\rm out})$ is the efficiency of energy generation,
\begin{equation}
\epsilon(R_{\rm in}, R_{\rm out}) = (e_{\rm out} - e_{\rm in}) - \Omega_{\rm out} (\ell_{\rm out} - \ell_{\rm in})  .
\label{efficiency-interior}
\end{equation}
For very large PDs, $R_{\rm out} \rightarrow \infty$, we have $e_{\rm out} = 0 = \Omega_{\rm out}$, and using the inner
boundary condition $\ell_{\rm in} = \ell_{\rm K}(R_{\rm in})$, we may write
\begin{equation}
\epsilon(R_{\rm in}, \infty) \equiv \epsilon_{\infty}(R_{\rm in}) = - e_{\rm K}(R_{\rm in}) = \frac{GM(R_{\rm in} - 2r_{\rm
G})}{2 (R_{\rm in} - r_{\mathrm S})^2}.
\label{efficiency-infinity}
\end{equation}
The efficiency $\epsilon_{\infty}(R_{\rm in})$ is the upper limit for the efficiency of a~PD with an inner radius $R_{\rm in}$. Nevertheless, we note that for $R_{\rm in} \rightarrow R_{\rm mb} = 2r_{\mathrm S}$ the efficiency of a~large PD tends to zero,
\begin{equation}
\epsilon_{\infty}(R_{\rm mb}) = 0.
\label{efficiency-zero}
\end{equation}
In thermal equilibrium the energy gain must be compensated for by the radiative and advective losses
\begin{equation}
({\rm total~heating}:~ L_{0})  = ({\rm total~cooling}:~ L + L_{\rm adv}).
\label{heating-cooling}
\end{equation}
\subsection{Polish doughnut advective cooling}
Since in the PD formalism no interior physics is considered, the advective losses cannot be calculated directly. We parameterize
the advective losses by $L_{\rm adv} = \xi L_{0}$, with some $0 \le \xi \le 1$. With this parametrization we write an 
explicit formula for the accretion rate in terms of luminosity,
\begin{equation}
{\dot M} = \frac{1}{\epsilon(1 - \xi)} L.
\label{accretion-rate}
\end{equation}
The total radiative efficiency of large PDs, that is, the upper limit of PD efficiency for given $R_{\rm in}$ and $\xi$, may be defined as
\begin{equation}
\epsilon_{\rm rad}(R_{\rm in}, \xi) = (1 - \xi) \epsilon_{\infty}(R_{\rm in}).
\label{total-rad-efficiency}
\end{equation}
Since we are interested in placing constraints on the thickness and luminosity for a~given  $\dot{m}$ of PDs, we use the highest efficiency $\epsilon_{\rm rad}$ in the calculations below. Radiatively inefficient accretion flows, or  RIAFs, have $\epsilon_{\rm rad}(R_{\rm in}, \xi) \ll 1$. The RIAF-type large PDs
have either $R_{\rm in} \approx R_{\rm mb}$, or $ \xi \approx 1$, or both.

\section{Extremal family of PDs}
\label{sec:maximalThick}
Models of PDs are determined by the specific angular momentum distribution. In this section we determine a~particular
distribution that for a~given location of the inner edge $R_{\rm in}$ gives the greatest possible relative thickness, $h = H/R$, 
and the highest possible efficiency of a~PD. 

Let $R_{\rm max}$ denote the (radial) location of the maximal relative thickness $h$,
for some unspecified angular momentum distribution
\begin{equation}
0 = \left(\frac{d h}{d R}\right)_{R_{\rm max}} = \frac{1}{R_{\rm max}} \left[ \left( \frac{d H}{d R}\right )_{R_{\rm max}} - \chi \right],
\label{eq:generalAngMom1}
\end{equation}
where $\chi = h(R_{\rm max})$ is the actual value of the greatest relative thickness, or in other words, the quantity that we wish to determine at its greatest extent. By ``funnel'' we understand the inner region of the disk, for which  $R < R_{\rm max}$, see Fig. \ref{Figure-constant-shape} (right).

From Eqs. (\ref{integration-shape}) and (\ref{eq:generalAngMom1}) we derive
\begin{equation}
\ell^2(R_{\rm max}) = \frac{GM R^3_{\rm max}(1 + \chi^2)^{1/2}}{\left[R_{\rm  max}(1 + \chi^2)^{1/2} - r_{\mathrm S} \right]^2} .
\label{eq:generalAngMom2}
\end{equation}
We introduce a parameter $y = \ell(R_{\rm max})/ \ell_{\rm K}(R_{\rm max}) $ that indicates how close the angular
momentum at $R_{\rm max}$ is to the local Keplerian value, and rewrite Eq. (\ref{eq:generalAngMom1}) as
\begin{equation}
a^2 - \left[ \frac{1}{y^2}(1 -\Delta)^2 + 2\Delta\right] a + \Delta^2 = 0,
\label{shape-solution}
\end{equation}
where $a^2 = 1 + \chi^2$ and $\Delta = r_{\mathrm S}/R_{\rm max}$. We note that
\begin{equation}
a = \frac{1}{y^2} + {\cal O}(\Delta),
\label{shape-solution-01}
\end{equation}
and therefore in the limit $\Delta \rightarrow 0$, that is,
\begin{equation}
R_{\rm max} \gg r_{\mathrm S},
\label{shape-solution-02}
\end{equation}
it is clear that the lower the angular momentum, the thicker the torus. Admissible specific angular momentum distributions
are non-decreasing in radius as a consequence of the Rayleigh stability condition. Hence, tori with angular momentum 
$\ell=\ell_K(R_{\rm in}) = \text{const}$ (at least for $R < R_{\rm max}$) represent a~family of extremely thick tori: the relative thickness $h$ is maximal for them. For these tori we find an analytic formula for $R_{\rm max}$,
\begin{equation}
R_{\rm max} = \left[\frac{R_{ \rm in}^3 r_{\mathrm S}}{(2R_{\rm in}-3 r_{\mathrm S})(R_{\rm in} - 2 r_{\mathrm S})} \right]^{1/2} 
\label{eq:Rmax}
\end{equation}
and, given Eq. (\ref{constant-shape}) and Eq. (\ref{eq:Rmax}), for the relative thickness,
\begin{equation}
\chi(R_{\rm in}) = \frac{H(R_{\rm max})}{R_{\rm max}} = h(R_{\rm max}).
\label{constant-maximal-ratio-function}
\end{equation}
The angular momentum
distribution for $R>R_{\rm max}$ is irrelevant for the maximization of $h$ -- it may be constant, but it may as well not be. In particular, any distribution that is constant for $R_{\rm in}\le R \le R_{\rm max}$ and has a~tail for $R_{\rm max} \le R \le R_{\rm out} $ chosen to satisfy Eq. (\ref{integral-condition}) for $R_{\rm out} \rightarrow \infty$, maximizes both geometrical thickness and radiative efficiency. Such tori constitute our family of extremal PDs, with the relative thickness given by Eq. (\ref{constant-maximal-ratio-function}) and efficiency given by Eq. (\ref{total-rad-efficiency}).
%
%
%
   \begin{figure}
   \centering
 \includegraphics[width=0.48\textwidth]{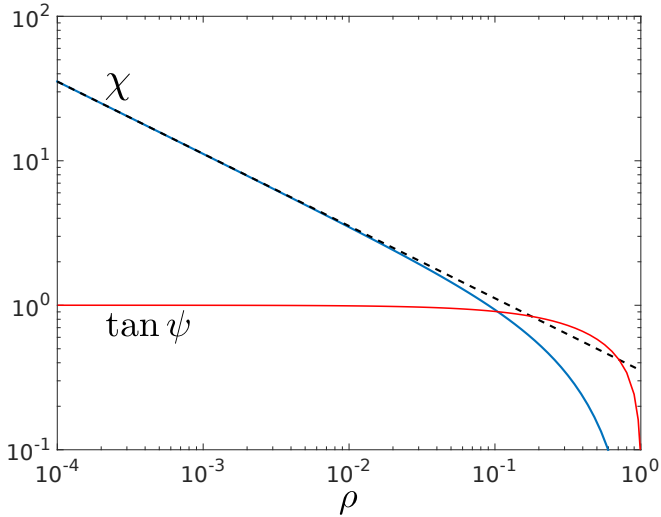}
   \caption{Highest value of the relative thickness $ \chi = (H/R)_{\rm max}$ for PDs. The straight dash-dotted line represents the asymptotic relation $\chi = 1/\sqrt{8\rho}$. The cusp opening half-angle $\psi$ is also indicated.}%
   \label{Figure-constant-H-R-max}
   \end{figure}
%

   \begin{figure}
   \centering
  \includegraphics[width=0.48\textwidth]{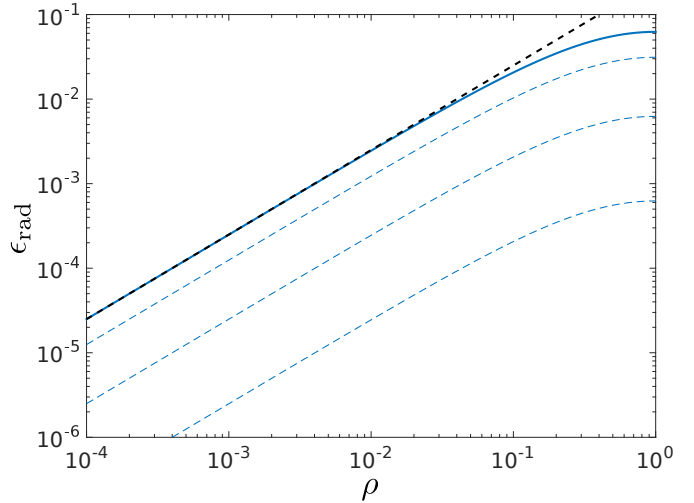}
   \caption{Radiative efficiency $\epsilon_{\rm rad}$. The continues line shows $\xi = 0$, the dashed lines represent advection strength $\xi = 0.5, 0.9,$ and $0.99$. The higher the advection, the lower the efficiency. The dash-dotted line represents the asymptotic formula for an efficiency $\eta = \rho/4$.}%
   \label{Figure-constant-efficiency}
   \end{figure}

   \begin{figure}   
      \centering
   \includegraphics[width=0.48\textwidth]{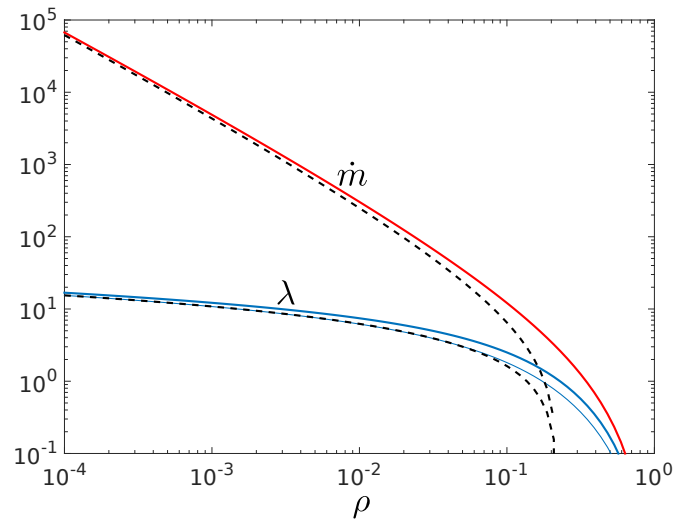}
   \caption{Luminosity $\lambda$ and accretion rate $\dot{m}$ for large, non-advective PDs with constant angular momentum. The luminosity integrated over whole torus is indicated with a~thick line, while the thin line corresponds to the funnel luminosity (i.e., integrated over $R < R_{\rm max}$). The dash-dotted lines correspond to the asymptotic formulae for very thick tori, Eqs. (\ref{eq:asymptotic_lambda} - \ref{eq:asymptotic_mdot}).  }%
   \label{Figure-constant-luminosity-and-accretion}
 \end{figure}
\section{Properties of very thick PDs}
\label{section-efficiency-limit}

In this section we derive the asymptotic expansion of the PD properties in the limit of very geometrically thick disks, $\chi \gg 1$. We expand the characteristic quantities up to a~leading term in the dimensionless parameter $\rho$, as defined by Eq. (\ref{eq:rho}), around $\rho = 0$. Immediately,
\begin{equation}
R_{\rm in} = (\rho + 2)r_{\mathrm S}.
\end{equation}
From Eq. (\ref{eq:Rmax}) we find
\begin{equation}
R_{\rm max} = \sqrt{ \frac{(\rho+2)^3}{\rho (2 \rho + 1)} }r_{\mathrm S} \approx \sqrt{\frac{8}{\rho}} r_{\mathrm S}.
\end{equation}
To find the expression for the geometric thickness we use Eq. (\ref{constant-shape}) to find, after some algebra,
\begin{equation}
1 + \chi^2  \approx \frac{1}{8\rho} ,
\label{R-max-argument-02}
\end{equation}
from which 
\begin{equation}
\chi \approx \sqrt{\frac{1}{8 \rho}}
\label{eq:asymptotic_chi}
\end{equation}
follows.
Hence, very thick tori indeed correspond to $\rho \rightarrow 0$, $R_{\rm in} \rightarrow R_{\rm mb}$, see Fig. \ref{Figure-constant-H-R-max}.
At the inner edge of the torus a~{\sl cusp} is formed, through which matter may be dynamically advected into the black hole. From Eq. (\ref{constant-shape}) we find that the cusp opening half-angle $\psi$ (see Fig. \ref{Figure-constant-shape}) obeys
\begin{equation}
\tan \psi = \sqrt{\frac{3 r_{\mathrm S} - R_{\rm in} }{R_{\rm in} - r_{\mathrm S}}}  = \sqrt{\frac{1-\rho}{1+\rho}} \approx 1- \rho.
\end{equation}
In the limit case of $\rho \rightarrow 0$ we find $\psi \rightarrow \pi/4$. As can be seen in Fig. \ref{Figure-constant-H-R-max}, the dependence of the cusp opening half-angle on $\rho$ is rather weak, and $\psi = \pi/4$ is a~good approximation for all very thick tori.

\begin{figure}
   \centering
   \includegraphics[width=0.47\textwidth]{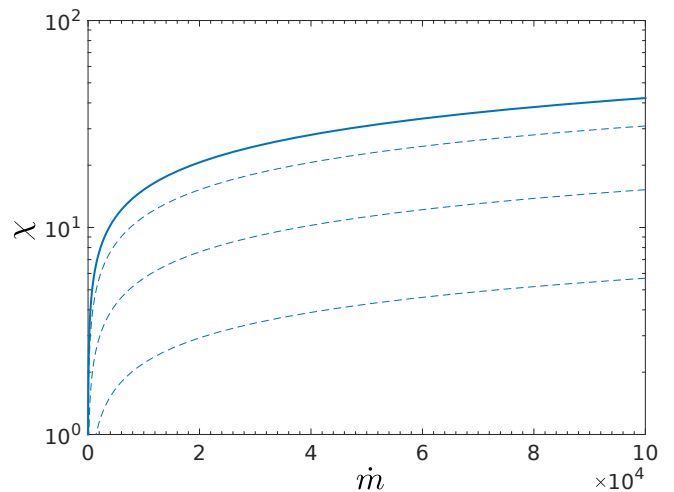}
   \caption{PD relative thickness $\chi$ as a function of $\dot{m}$. The continuous line shows $\xi = 0$, the dashed lines represent $\xi = 0.5, 0.9,$ and $0.99$.
}
   \label{Figure-thickness-mdot}
 \end{figure}

The limit efficiency, Eq. (\ref{total-rad-efficiency}), can be expanded as
\begin{equation}
\epsilon_{\rm rad} = \frac{GM}{2} \frac{R_{\rm in}-2r_{\mathrm S}}{(R_{\rm in}-r_{\mathrm S})^2}(1 - \xi) \approx \frac{\rho}{4} (1 - \xi).
\end{equation}
This is shown in Fig. \ref{Figure-constant-efficiency}. Since $\tan \alpha = 1/\chi \approx \sqrt{8 \rho}$, the funnel collimating factor is
\begin{equation}
\beta  = (1 - \cos \alpha)^{-1} = \left[ 1 - \left(1 + 8 \rho  \right)^{-1/2} \right]^{-1} \approx \frac{1}{4 \rho} \ .
\end{equation}
Expanding the luminosity is much more involved because we need to integrate over a~non-trivial PD surface. However, assuming a~conical shape of the funnel and integrating the radiation flux over such a~surface (we recall that for very thick tori $L_{\rm fun} \approx L$), we find the following result
\begin{align}
& \lambda \approx \frac{L_{\rm fun}}{ L_{\rm Edd}} =  2 \cos \alpha \int^{R_{\rm max}}_{R_{\rm in}} \frac {R dR}{(R - r_{\mathrm S})^2} =  \\
& 2 \cos \alpha  \ln \left(\frac{R_{\rm max} - r_{\mathrm S}}{R_{\rm in} - r_{\mathrm S}}\right) + \frac{2  r_{\mathrm S} (R_{\rm max} - R_{\rm in}) \cos \alpha}{(R_{\rm max} - r_{\mathrm S})(R_{\rm
in} - r_{\mathrm S})}, \nonumber
\label{luminosity-cone-PW}
\end{align}
which can be expanded for $\rho \approx 0$ as
\begin{equation}
\lambda \approx \frac{L_{\rm fun}}{L_{\rm Edd}} \approx \cos \alpha \ln \left[ \frac{8 \exp(2)}{\rho} \right ] .
\label{eq:frustumLum}
\end{equation}
Formula (\ref{eq:frustumLum}) is not expected to yield an accurate result. We find that it estimates the luminosity with a~relative error of about 50\%. We proceed by assuming that Eq. (\ref{eq:frustumLum}) at least provides a~proper functional form of the asymptotic relation between $\rho$ and $L$ in the form
\begin{equation}
\lambda \approx A \ln \frac{B}{\rho}
\label{eq:asymptotic_lambda}
\end{equation}
for some constants $A$ and $B$. An accurate fit can be found for $A = 2$ and $B = 0.225$, see Fig. \ref{Figure-constant-luminosity-and-accretion}.
Combining the results for luminosity and efficiency, we find the mass accretion rate
\begin{equation}
\dot{m} = \frac{0.1 \lambda}{\epsilon_{\rm rad}} \approx \frac{0.8}{\rho (1 - \xi)} \ln \left( \frac{0.225}{\rho} \right) .
\label{eq:asymptotic_mdot}
\end{equation}
We see now that because $\epsilon_{\rm rad} \sim \rho \sim 1/ \beta $, the collimated luminosity is
\begin{equation}
L_{\rm col} = \beta L_{\rm fun} \approx \beta \epsilon_{\rm rad} \dot{M}  \approx 0.0625 \dot{M}(1-\xi) \ ,
\end{equation}
or, in dimensionless units,
\begin{equation}
\lambda_{\rm col} = \beta \lambda_{\rm fun} \approx \beta \epsilon_{\rm rad} 10 \dot{m} \approx 0.625 \dot{m} (1-\xi) \ .
\end{equation}
Clearly, for very thick PDs, the luminosity collimated by the narrow funnel scales linearly with the mass accretion rate.
From plotting the relative thickness $\chi$ against the mass accretion rate $\dot{m}$, Fig. \ref{Figure-thickness-mdot}, we see that while formally the PDs can be arbitrarily thick, even a~huge accretion rate of $10^5 \dot{M}_{\rm Edd}$ only provides $\chi \approx 40$. Furthermore, strong advection reduces the thickness by a~factor on the order of $(1 - \xi)^{1/2}$, and for an advection parameter $\xi = 0.99$ even  $\dot{M} = 10^5 \dot{M}_{\rm Edd}$ cannot produce a~disk with a~relative thickness larger than $\chi = 6$.


\subsection{Accretion ``branches''}
\label{subsection-efficiency-limit}

\begin{figure}
   \centering
   \includegraphics[width=0.46\textwidth, trim={0mm 0 0 0},clip]{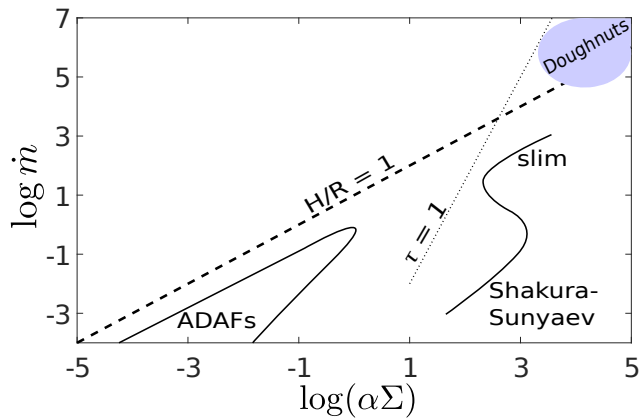}
   \caption{ Accretion disk equilibria at a particular radial location $R = 5 r_{\mathrm S}$ and for a~black hole mass $10 M_\odot$. Flows above the dotted lines $\tau = 1$ are optically thin. The solid S-shaped line on the right represents the optically
thick disks, the solid line on the left represents the
optically thin disks. The upper branches of both lines represent
advection-dominated disks (slim and ADAFs).The Shakura-Sunyaev viscosity parameter is assumed to be $\alpha = 0.1$. We note the
location of the PDs. This figure is adopted (with some simplifications) from \cite{LasotaIntro}, the original appeared in \cite{Chen-1995}.
}
   \label{Figure-branches}
 \end{figure}

It is customary to display analytic models of accretion disks in the ${\dot m}$ versus $\Sigma(R)$ parameter space, where
\begin{equation}
\Sigma(R) = \int^{+H(R)}_{-H(R)} \rho(R, Z) dZ,
\label{surface-density}
\end{equation}
is the surface density at a given cylindrical location $R$. An example is shown in Fig. \ref{Figure-branches} which locates the
main types of thin (or semi-thin) disk models: Shakura-Sunyaev, slim, and ADAFs. The question of where in this figure the PDs are located, cannot be answered precisely and unambiguously. There is no unique model for the PD interiors, therefore the function $\Sigma(R)$ is not known.
In the ``classic'' zero-advection case considered in the 1980s , the region of the thick PDs is above
the $H/R = 1$ line and to the right of the $\tau = 1$ line, as for example in the review by \cite{Abramowicz-2013}. However, when
strong advection is added, as we did here, the PDs move into the region
occupied by slim disks.
\section{Conclusions}
We have demonstrated that at high accretion rates, when advection probably is a~dominant source of cooling, the relative thickness
of PDs is significantly reduced. Doughnuts with strong advection may be considered as approximate models of slim
disks, which is consistent with the conclusion of \citet{Oleketal15}. We considered advection as an additional (to radiation) cooling process, using only very general, global,
conservation laws for mass, energy, and angular momentum. In particular, we did not specify whether advection is radial or
vertical. Thus it follows from our general considerations that any type of strong advection would keep the thickness of the PDs at relatively low values. Our results cast doubts on whether collimation by a thick-disk funnel is an adequate model for the ULXs. When advection is taken into account, even very high mass accretion rates cannot produce sufficient collimated luminosity, cf. Fig. \ref{Figure-main-results}. Moreover, while a~non-advective disk seems to agree with the numerical GRMHD results of \cite{Sadowski-2015}, the latter does not report any increase of thickness with the mass accretion rate.

To conclude, PDs are very ``minimalist'' models of accretion flows at super-Eddigton accretion rates. They only give the  ``photospheric'' properties of these flows and do not refer to their interiors. This is  both a~deficiency and a~virtue.


\begin{acknowledgements}
Discussions on super-Eddington flows with Andrew King and Olek S\c adowski were of great help when writing this paper. This work was supported by the Polish NCN grants 2013/09/B/ST9/00060,  DEC-2012/04/A/ST9/00083  and UMO-2013/08/A/ST9/00795, the Czech ``Synergy'' (Opava) grant ASCRM100031242 CZ.1.07/2.3.00/20.0071. MW acknowledges support of the Foundation for Polish Science within  the START Programme. JPL was supported in part by the French Space Agency CNES.
\end{acknowledgements}
%
%
\bibliography{biblio}
\bibliographystyle{aa}

\end{document}